\documentclass{PoS}

\title{Study of hadron and gamma-ray acceptance of the MAGIC telescopes: towards an improved background estimation}

\ShortTitle{Acceptance of the MAGIC telescopes}

\author{\speaker{Elisa Prandini}\\ 
        Astronomy Department, University of Geneva\\
        E-mail: \email{elisa.prandini@unige.ch}}
\author{ Giovanna Pedaletti\\
        DESY, Zeuthen}
\author{Paolo Da Vela \\
        INFN Pisa}
\author{Emma de Ona Wilhelmi \\
        Institute of Space Sciences CSIC/IEEC, Barcelona}
\author{Pierre Colin, Christian Fruck,  Marcel Strzys, Ievgen Vovk\\
        MPI, Munich}
\author{for the MAGIC Collaboration}

\abstract{The MAGIC telescopes are an array of two imaging atmospheric Cherenkov telescopes (IACTs) studying the gamma ray sky at very high-energies (VHE; E>100 GeV). The observations are performed in stereoscopic mode, with both telescopes pointing at the same position in the sky. The MAGIC field of view (FoV) acceptance for hadrons and gamma rays has a complex shape, which depends on several parameters such as the azimuth and zenith angle of the observations. In the standard MAGIC analysis, the strategy adopted for estimating this acceptance is not optimal in the case of complex FoVs.

In this contribution we present the results of systematic studies intended to characterise the acceptance for the entire FoV. These studies open the possibility to apply improved background estimation methods to the MAGIC data, useful to investigate the morphology of extended or multiple sources.}

\FullConference{The 34th International Cosmic Ray Conference,\\
		30 July- 6 August, 2015\\
		The Hague, The Netherlands}

\begin{document}

\section{Introduction}
Imaging Atmospheric Cherenkov Telescopes (IACTs) are ground-based telescopes that study the gamma ray sky at very high-energies (VHE; E>100 GeV).
Current generation of IACTs include the successful experiments H.E.S.S., located in Namibia, MAGIC, in the Canary Islands, and VERITAS, in Arizona.

Due to the opacity of the atmosphere to gamma rays, IACTs observe a by-product of electromagnetic cascades initiated in the atmosphere by very energetic gamma-rays, namely the Cherenkov light. This light is focused into each telescope camera and forms an image that is analyzed in order to characterize the primary particle and determine its energy and incoming direction. 

The {\it camera acceptance} can be defined as the  systemic response for the detection of $\gamma$-ray events \cite{fernandes14}. When observing with a single telescope, the camera acceptance is in first approximation radially symmetric, with the maximum at the camera center, meaning that the system is more sensitive to light emitted by gamma-like showers focused in the central part of the camera than in the outer part.  Second order asymmetries are due to large zenith angle observations and also to the geomagnetic field.
With two or more telescopes, the camera acceptance becomes more complicated, being the intersection of each  trigger telescope field of view (FoV). The camera acceptance is  expected to be different for VHE gamma-ray induced images and background images, typically due to proton-induced showers. It  can also depends on other parameters \cite{hess_template}. 

We present a detailed study of the camera acceptance of the MAGIC telescopes. We will discuss its dependence on the observation parameters (such as azimuth and zenith angle) and on the nature of the particle initiating the atmospheric shower.

This study is motivated by the possibility of applying new sophisticated methods for the background reconstruction, such as the template background method \cite{hess_template}. This method is suited for the reconstruction of sky maps of complex FoVs, and is based on the detailed knowledge of both gamma and hadron-like event acceptances.

\section{The MAGIC Telescopes}
MAGIC is a system of two IACTs located in the Canary Island of La Palma, at 2200~m asl. 
It observes VHE gamma rays from 50~GeV up to several tens of TeV. 
The angular resolution at around 200~GeV energies is < 0.07 degree, while the energy resolution is 16\% \cite{mag_analysis, magicperf14}.

During the observation of a VHE gamma-ray candidate, 
the system records the azimuth (Az) and zenith (Zd) angle of each observation.
The system acceptance is the intersection of the two radially symmetric telescopes acceptances, and in first approximation is an ellipse centered in the camera center, at a given azimuth and zenith angle.

In order to characterize systematically the MAGIC camera acceptance, we studied a large sample of real data collected by the telescopes between the end of 2013 and Summer 2014. 
 
\subsection{Data selection and analysis chain}
\begin{figure}[htp]
  \centering
  \includegraphics[width=3.5in]{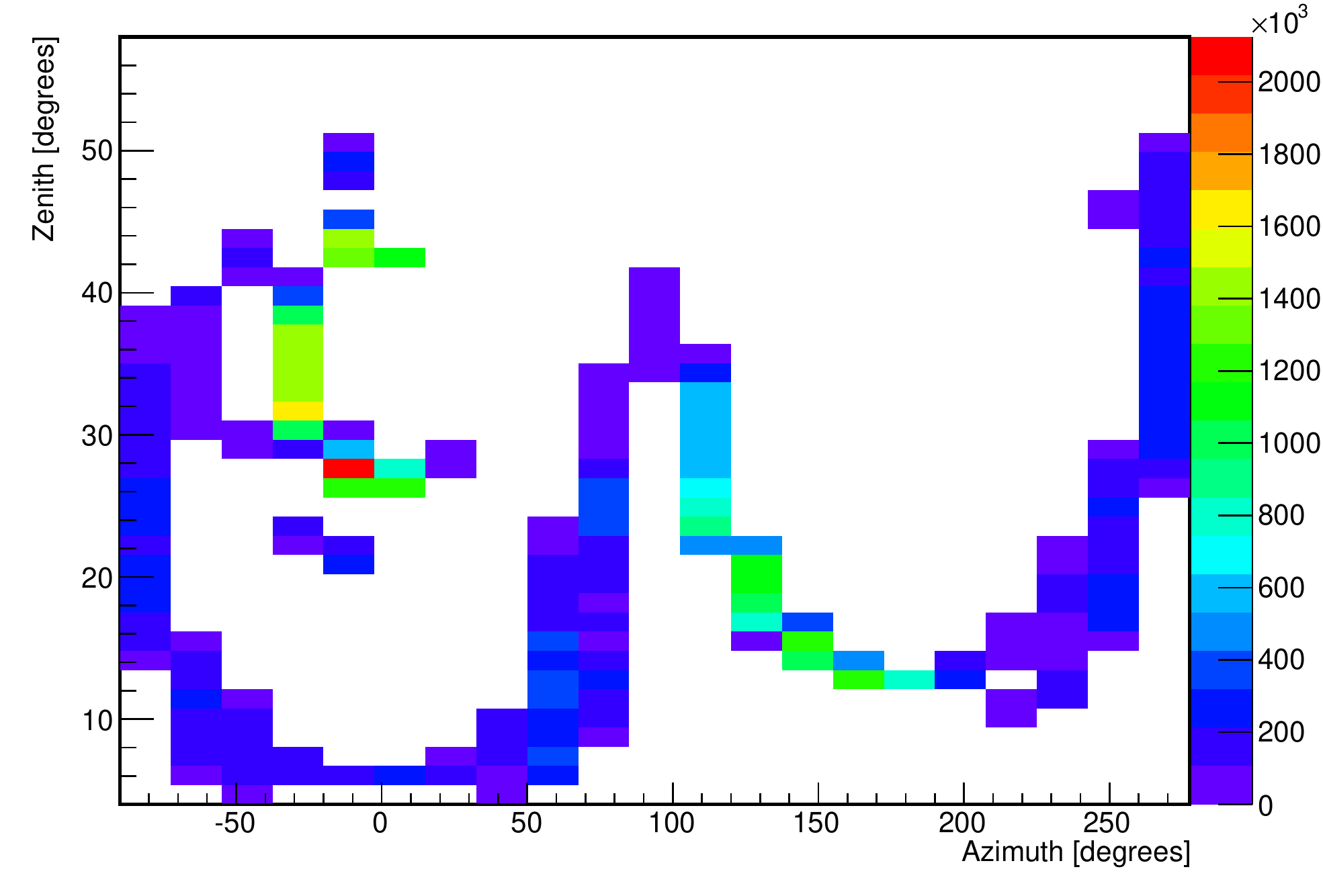}
  \caption{Zenith versus azimuth angle distribution for the 79 hours of data ($5.9~\cdot~10^7$ events) used in this study. For graphical reasons, the azimuth is displayed in the range -80 to 280 degrees (instead of 0 to 360 degrees).}
  \label{fig:data_azzd}
 \end{figure}

For the study of the MAGIC acceptance, we selected 79~hours of so-called Off data, i.e. data without any significant gamma-ray signal. The good Az-Zd coverage of this dataset is shown in Figure~\ref{fig:data_azzd}.
The data selection was based mainly on the rate of the events, an indicator of good weather conditions and dark sky. The MAGIC telescopes can observe also during low and moderate moonlight conditions. For the dataset used here, we allow only data taken during a low level of moonlight for which the analysis pipeline is not altered with respect to dark sky conditions. The data were processed using the standard MAGIC data analysis chain \cite{mag_analysis}.

In the analysis chain, a geometrical reconstruction of the camera image is done for each triggered event, following the extended Hillas parametrization, \cite{hillas85}. From this reconstruction, the energy and nature of the primary particle as well as its reconstructed arrival direction is then estimated. Even for strong gamma-ray emitters, such as the Crab Nebula, the large majority of these images are associated with hadron initiated showers.

In order to provide an independent check of the reported results, a second analysis was done  with a partially different dataset both for signal and background characterization. In this analysis, we have found results compatible with those reported in the following. 

\section{Study of the Acceptance}
\begin{figure}[htp]
  \centering
  \includegraphics[width=2.5in]{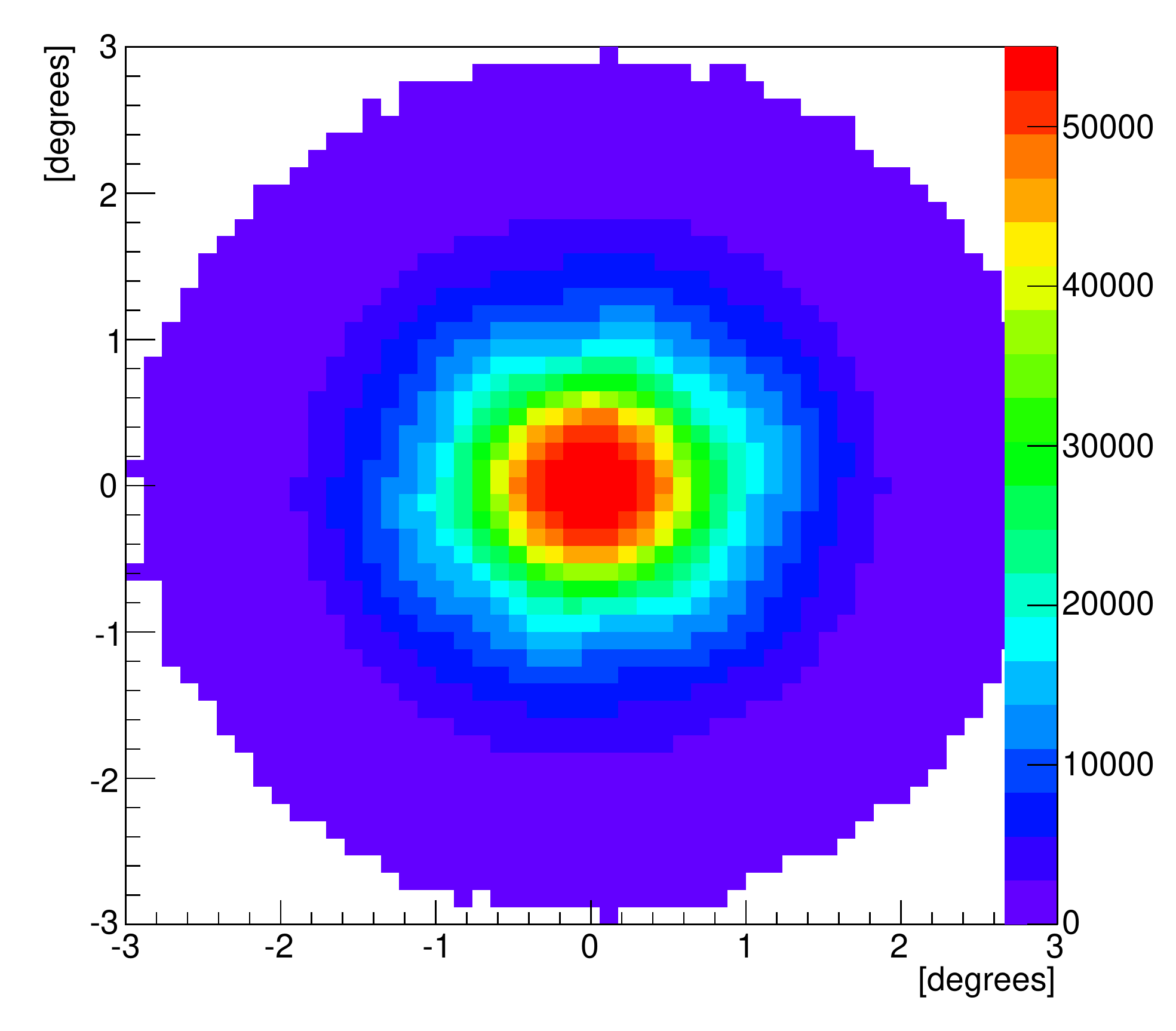}
  \caption{Acceptance map for the 79~hours of Off data ($5.9~\cdot~10^6$ events) considered in this study.}
  \label{fig:acc_all}
 \end{figure}

The arrival direction  estimation is refined during the last step of the analysis chain with the random forest technique \cite{mag_analysis}, where the random forest is trained on simulated gamma rays. For pure geometrical reasons, this incoming direction is a point in the camera.
The plot in camera coordinates (in degrees) of the reconstructed incoming directions is used to describe the acceptance of the MAGIC telescopes, Figure~\ref{fig:acc_all}. The binning adopted in the study is 51 bins from -3 to 3 degrees both for X and Y axis.

\subsection{Azimuthal dependence of the acceptance}
\begin{figure}[htp]
  \centering
  \includegraphics[width=\textwidth]{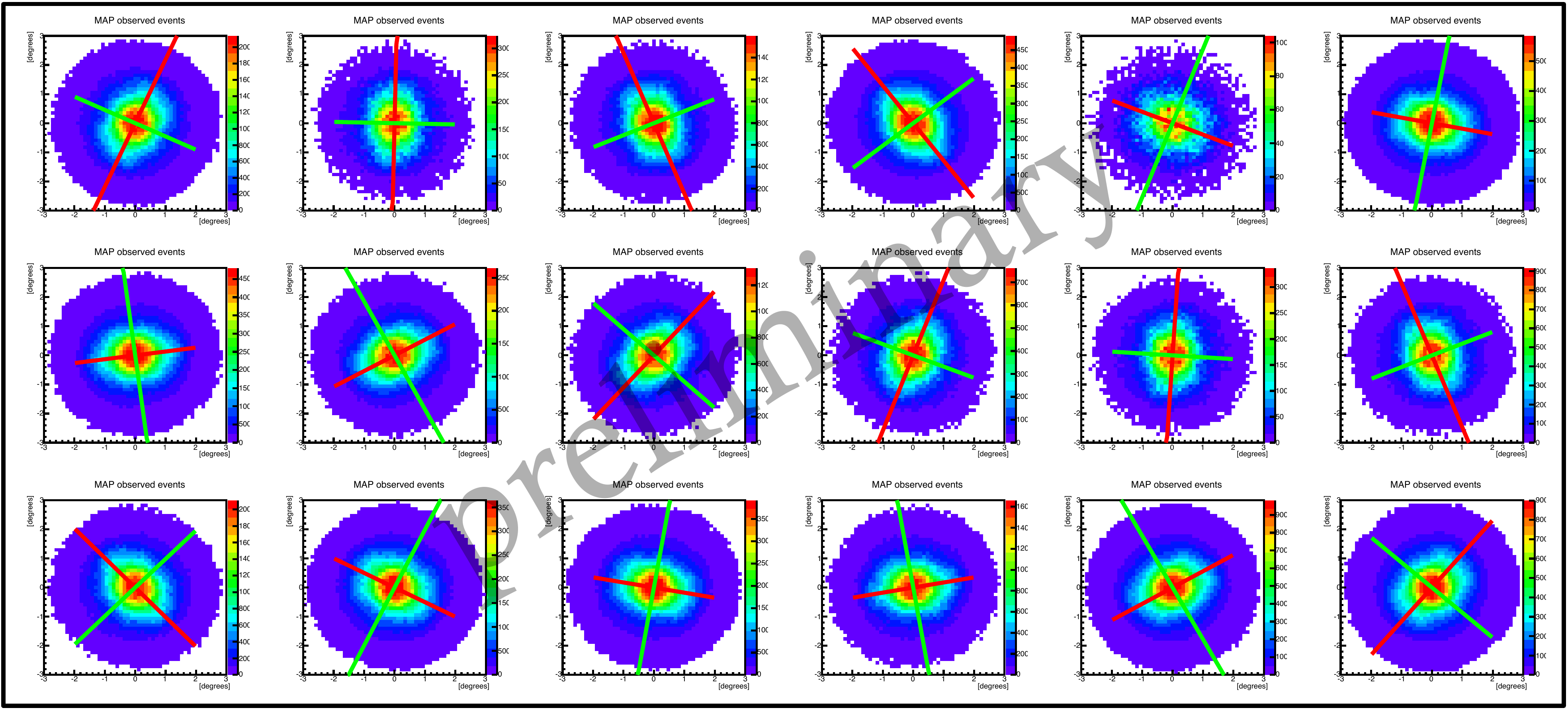}
  \caption{From left to right, top to bottom: acceptance map in 18 azimuth bins, from 0 to 360 degrees, in step of 20 degrees. The first bin (azimuth 0-20 degrees) is drawn in the upper row on the left, while the last bin (azimuth 340-360 degrees) is drawn in the bottom row on the right.}
  \label{fig:acc_az}
 \end{figure}

In MAGIC data analysis, azimuth is counted from Geographic North (0 deg) in the direction of East (90 deg).
In order to characterize the effect of the azimuth angle of the observation on the acceptance, we divided our sample in 18 azimuth bins, spanning the full range in azimuth. 
Figure~\ref{fig:acc_az} shows the resulting acceptance plots, for each azimuth bin. 
It is worth noticing that, while in Figure~\ref{fig:acc_all} the acceptance shows mostly a radial dependence from the center of the camera, when we plot the acceptance in azimuth bin this dependance is complicated and the shape becomes elliptical, but still centered in the camera center. This effect is due to the interception of the acceptances of each MAGIC telescope, which is azimuth dependent\footnote{In the standard MAGIC analysis, this azimuth angle dependence is taken into account both in the spectral and morphological analyses, which are usually performed in bins of azimuth angle. However, the strategy adopted for estimating the background is not optimal in the case of complex FoVs.}.
In Figure~\ref{fig:acc_all} the effect is not visible because a large amount of data covering almost all the azimuth  range is considered.

The elliptic shape assumed by the acceptance in each azimuth bin can be parametrized by the eigenvectors of
the covariance matrix, which represent the directions of the semi-axes of the ellipse 
(red and green lines drawn in Fig.\ref{fig:acc_az}). The eigenvalues of the covariance matrix, 
instead, are related to the variance of the data, and are therefore connected to the shape of the acceptance.

\begin{figure}[htp]
  \centering
  \includegraphics[width=5.0in]{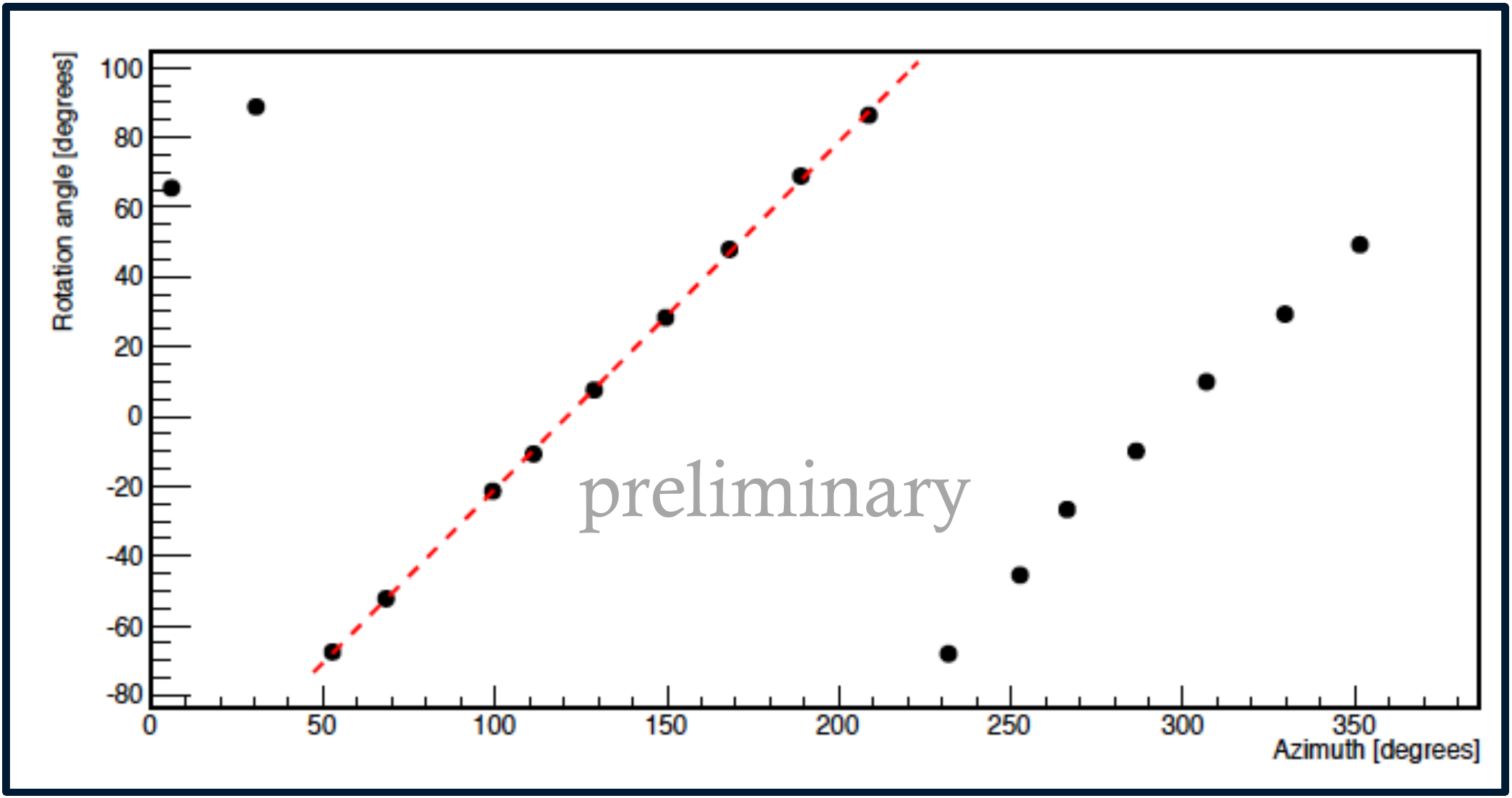}
  \caption{ Angle between the major axis of the ellipse describing the camera acceptance and the horizontal axis of the camera plane, as a function of the azimuth angle.} %Histogram representing the angle between the ellipse major axis and the horizontal axis of the camera plane, as a function of the azimuth angle.
  \label{fig:acc_angle}
 \end{figure}

Figure~\ref{fig:acc_angle} shows the dependance of the angle between the major axis of the ellipse and the 
horizontal axis of the camera plane with the azimuth angle of the observation. Its is evident that this relation is linear, and therefore, on first order, the azimuth dependance can be corrected by derotating the acceptance map.

Hence, to correct for the azimuth effect on the acceptance, the arrival direction coordinates of each event taken at a particular azimuth have to be 
derotated of an angle given by the formula:

\begin{equation}
\phi_0 = \phi - 120,
\end{equation}
where $\phi$ is the azimuth of the observation, and $\phi_0$ is the derotation angle, in degrees. This relation has been obtained by fitting the data shown in \ref{fig:acc_angle} in the azimuth range 40-240 degrees.

\subsection{Zenithal dependence of the acceptance}
\begin{figure}[htp]
  \centering
  \includegraphics[width=5.0in]{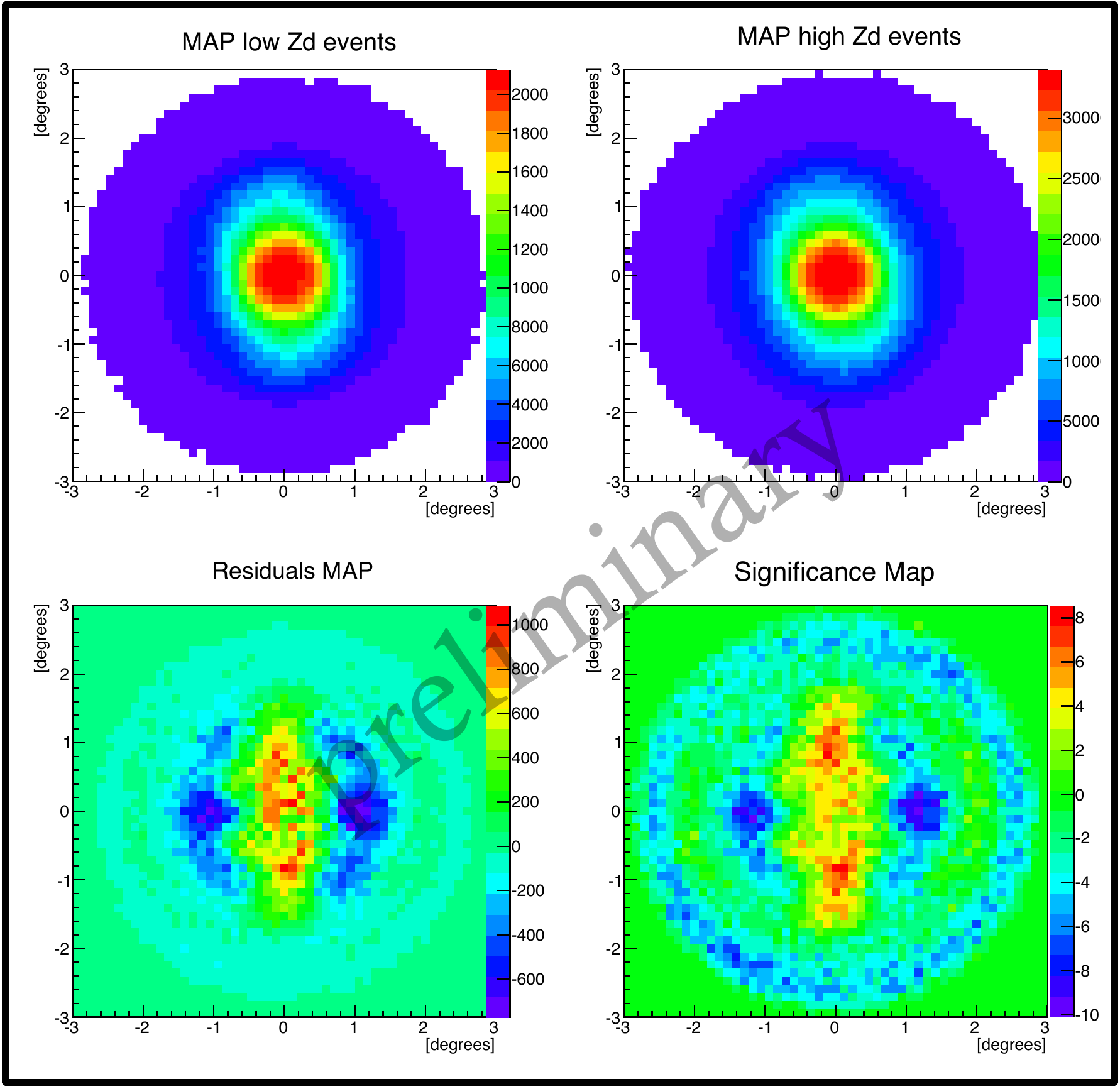}
  \caption {Upper panels: Acceptance of low and high zenith angles events (corrected for the azimuth dependence). Bottom panels: difference between the two normalized acceptances and corresponding Li\&Ma significance map.}
  \label{fig:acc_zd}
 \end{figure}

The comparison of low (< 22.5 degrees) and high (> 22.5 degrees) zenith angle acceptance maps for events with estimated energy larger than 250\,GeV is plotted in the first row of Figure~\ref{fig:acc_zd}.
To avoid contamination of the azimuth angle dependance, we have derotated the maps following the method described in the previous section. By visual inspection, it seems that the MAGIC telescopes acceptance for high zenith angle events is more roundish than that obtained for low zenith angle events.

This is confirmed by the study of the difference of the two maps, normalized for the total number of entries, and its significance obtained with the Li\&Ma formula \cite{lima}. Both graphs are drawn in the bottom row of Figure~\ref{fig:acc_zd}. 
The low zenith angle acceptance is found to be significantly more peaked at the camera center, and the ellipse describing this average acceptance is more elongated with respect to the one describing the acceptance to the high zenith angle events. A similar effect is observed when comparing events with low and high estimated energy.

In conclusion, we demonstrated that the zenith angle of the observation affects the acceptance of the MAGIC telescopes, and therefore it should be taken into account when the background is modeled.

\subsection{Comparison of hadron-like and gamma-like events acceptance}
A crucial point to be addressed when a IACT acceptance is studied, 
is the systemic response to gamma and background-induced (i.e. proton) showers.
As detailed in \cite{hess_template}, this is of particular relevance for the development of advanced image tools using for example the template background technique.

In order to study the MAGIC response to the two kind of showers, we compare the acceptance 
of our system to gamma-like and hadron-like events with estimated energy larger than 250\,GeV. 
During the data analysis, a parameter named {\it hadronness}  is assigned to each event. 
This parameter is related to the probability that a given event is 
induced or not by a primary gamma, and spans from 0 (gamma-like event) to 1 (hadron-like event). For the computation of the hadronness, the random forest technique is employed, where the random forest is trained on simulated gamma rays as well as on real hadronic showers.

In this study we have applied the following cuts:
\begin{itemize}
\item hadronness < 0.28: gamma-like events; 
\item 0.4 < hadronness < 0.8: hadron-like events;
\item size > 200 photo-electrons to all the events.
\end{itemize}

\begin{figure}[htp]
  \centering
  \includegraphics[width=5.0in]{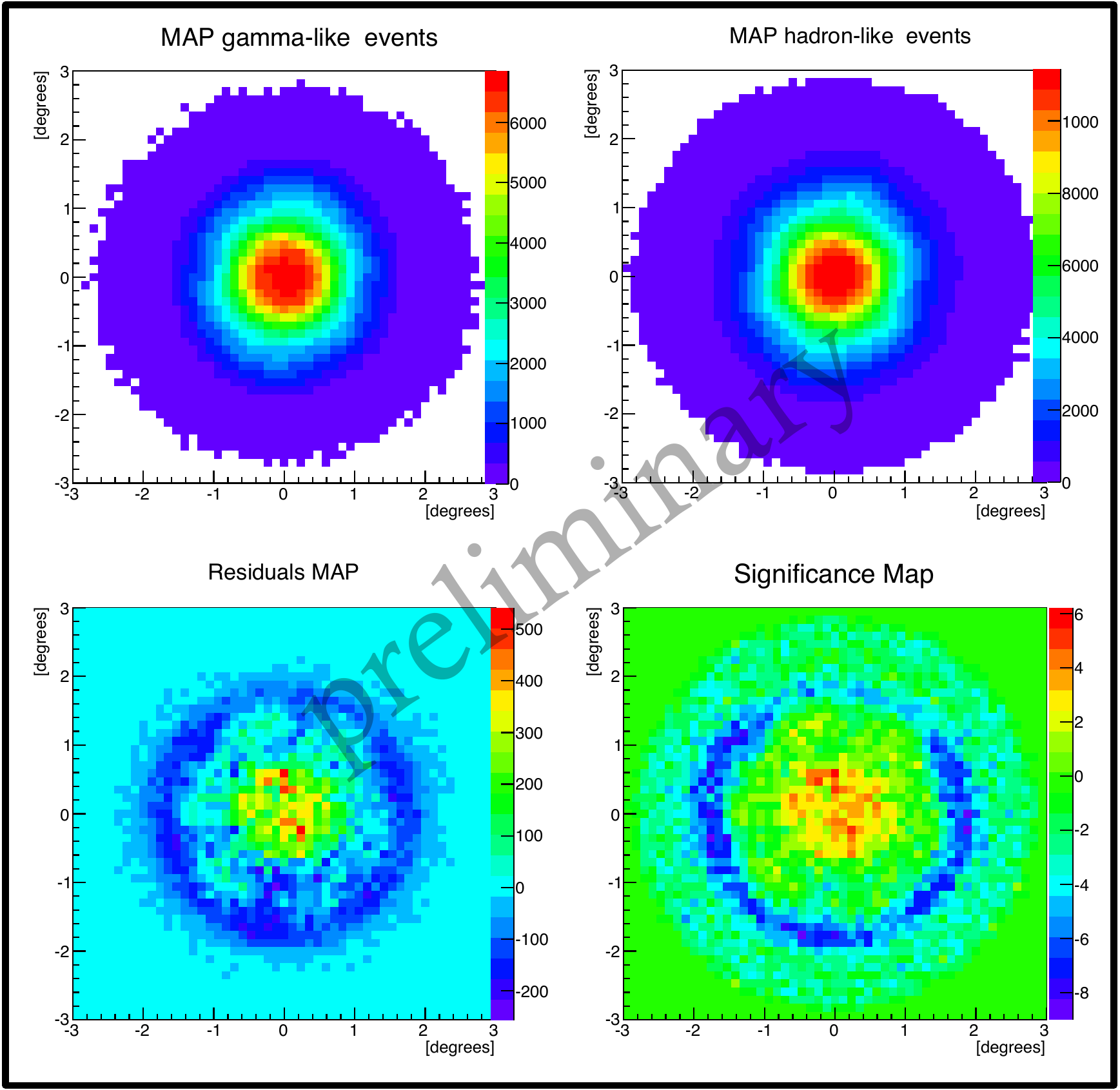}
  \caption{Upper panels: Acceptances of gamma- and hadron-like events. Bottom panels: difference between the two normalized acceptances, and Li\&Ma significance of the difference.}
  \label{fig:acc_hadronness}
 \end{figure}

The resulting acceptances are plotted in Figure~\ref{fig:acc_hadronness}. 
The first plot on the left of the upper panel represents the MAGIC 
acceptance to gamma-like showers, while the second plot is the acceptance for hadron-like showers. 
The difference between the two maps (normalized to the entries in the inner part of the camera) 
and its Li \& Ma significance is displayed on the bottom row.
These plots shows that there is a clear difference in the camera acceptances between gamma and hadron-like events.
In particular, the positive excesses at low radii and negative excesses at large radii show that the acceptance to gamma-like events is more peaked than the corresponding acceptance to hadron-like events.

\section{Conclusions}
In this proceedings, we have presented a detailed study of the MAGIC telescopes acceptance, 
based on the analysis of nearly 80 hours of real Off data. 
For a given observation, the acceptance has been found to have an elliptic shape. 
We have demonstrated that, as expected for purely geometrical reasons, different azimuth 
angles induce a rotation of this ellipse in the camera plane.

The zenith angle of the observations has also an effect on the MAGIC acceptance, which is more
roundish for large zenith angle observations.  

Finally, we have compared the hadron-like and gamma-like events acceptances, finding significant differences 
in the two responses. 

The characterization of the camera acceptance  of the MAGIC telescopes performed in this study
 opens the possibility of applying  improved background estimation methods to the MAGIC data, useful to investigate the morphology of extended or multiple sources.

\section*{Acknowledgements}
We would like to thank the Instituto de Astrof\'{\i}sica de Canarias
for the excellent working conditions at the Observatorio del Roque de los Muchachos in La Palma. The financial support of the German BMBF and MPG, the Italian INFN and INAF, the Swiss National Fund SNF, the ERDF under the Spanish MINECO (FPA2012-39502), and the Japanese JSPS and MEXT is gratefully acknowledged. This work was also supported by the Centro de Excelencia Severo Ochoa SEV-2012-0234, CPAN CSD2007-00042, and MultiDark CSD2009-00064 projects of the Spanish Consolider-Ingenio 2010 programme, by grant 268740 of the Academy of Finland, by the Croatian Science Foundation (HrZZ) Project 09/176 and the University of Rijeka Project 13.12.1.3.02, by the DFG Collaborative Research Centers SFB823/C4 and SFB876/C3, and by the Polish MNiSzW grant 745/N-HESS-MAGIC/2010/0.
Elisa Prandini gratefully acknowledges the financial support of the Marie Heim-Vogtlin grant of the Swiss National Science Foundation.

\end{document}